\documentclass[a4paper]{article}

\usepackage{INTERSPEECH2020}
\usepackage{hyperref}

\title{Are you wearing a mask? Improving mask detection from speech using augmentation by cycle-consistent GANs}
\name{Nicolae-C\u{a}t\u{a}lin Ristea$^1$, Radu Tudor Ionescu$^2$}
\address{
  $^1$University Politehnica of Bucharest, Romania\\
  $^2$University of Bucharest, Romania}
\email{r.catalin196@yahoo.ro, raducu.ionescu@gmail.com}

\begin{document}

\maketitle

\begin{abstract}
The task of detecting whether a person wears a face mask from speech is useful in modelling speech in forensic investigations, communication between surgeons or people protecting themselves against infectious diseases such as COVID-19. In this paper, we propose a novel data augmentation approach for mask detection from speech. Our approach is based on $(i)$ training Generative Adversarial Networks (GANs) with cycle-consistency loss to translate unpaired utterances between two classes (with mask and without mask), and on $(ii)$ generating new training utterances using the cycle-consistent GANs, assigning opposite labels to each translated utterance. Original and translated utterances are converted into spectrograms which are provided as input to a set of ResNet neural networks with various depths. The networks are combined into an ensemble through a Support Vector Machines (SVM) classifier. With this system, we participated in the Mask Sub-Challenge (MSC) of the INTERSPEECH 2020 Computational Paralinguistics Challenge, surpassing the baseline proposed by the organizers by 2.8\%. Our data augmentation technique provided a performance boost of 0.9\% on the private test set. Furthermore, we show that our data augmentation approach yields better results than other baseline and state-of-the-art augmentation methods.
\end{abstract}

\setlength{\abovedisplayskip}{3pt}
\setlength{\belowdisplayskip}{3pt}

\section{Introduction}

In this paper, we describe our system for the Mask Sub-Challenge (MSC) of the INTERSPEECH 2020 Computational Paralinguistics Challenge (ComParE) \cite{Schuller-INTERSPEECH-2020}. In MSC, the task is to determine if an utterance belongs to a person wearing a face mask or not. As noted by Schuller et al.~\cite{Schuller-INTERSPEECH-2020}, the task of detecting whether a speaker wears a face mask is useful in modelling speech in forensics or communication between surgeons. In the context of the COVID-19 pandemic, another potential application is to verify if people wear surgical masks.

We propose a system based on Support Vector Machines (SVM) \cite{Cortes-ML-1995} applied on top of feature embeddings concatenated from multiple ResNet \cite{He-CVPR-2016} convolutional neural networks (CNNs). In order to improve our mask detection performance, we propose a novel data augmentation technique that is aimed at eliminating biases in the training data distribution. Our data augmentation method is based on $(i)$ training Generative Adversarial Networks (GANs) with cycle-consistency loss \cite{Zhu-ICCV-2017,Kim-ICLR-2020} for unpaired utterance-to-utterance translation among two classes (with mask and without mask), and on $(ii)$ generating new training utterances using the cycle-consistent GANs, assigning opposite labels to each translated utterance.

\begin{figure*}[!t]
\begin{center}
\centerline{\includegraphics[width=0.9\linewidth]{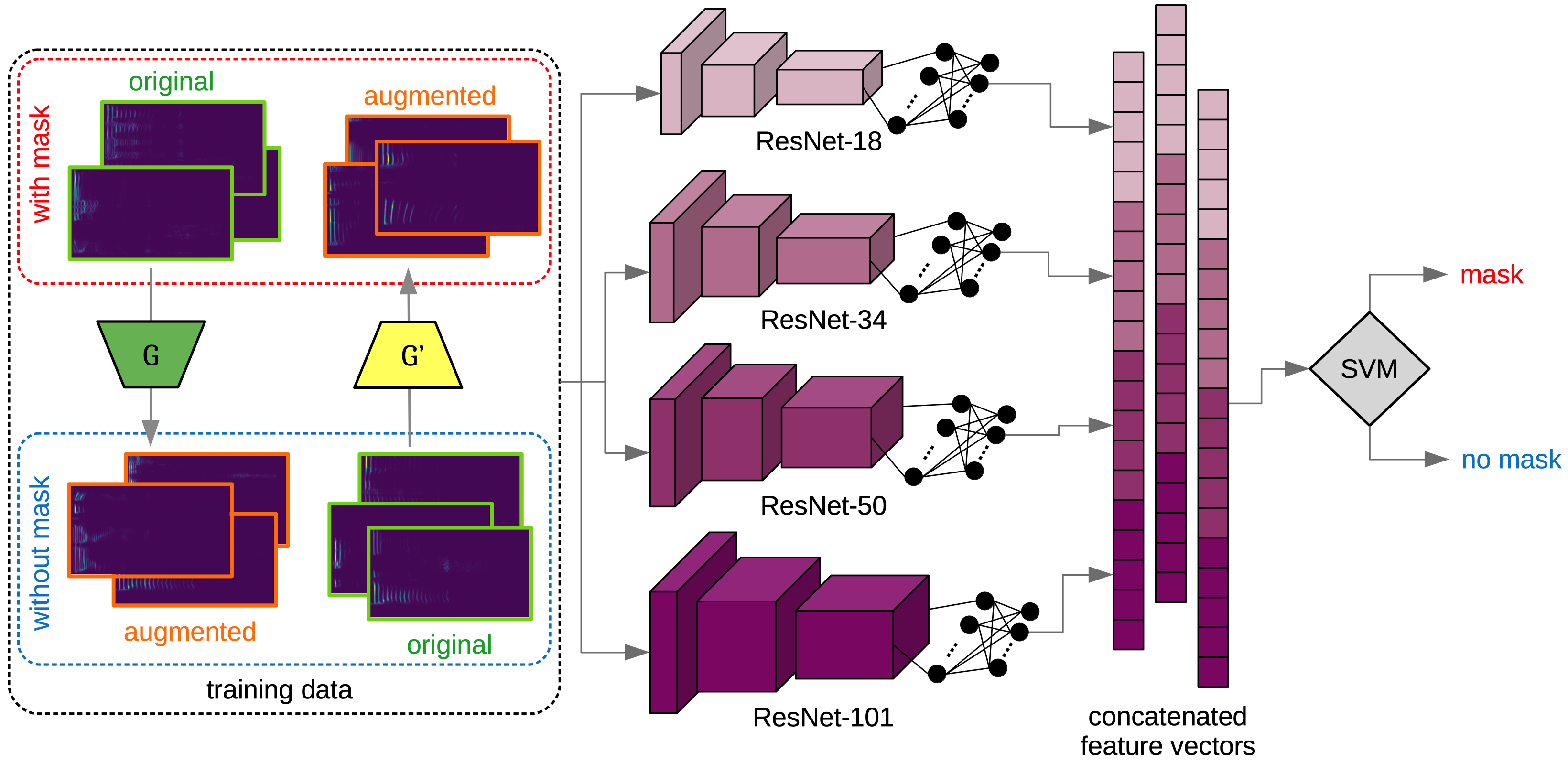}}
\vskip -0.2cm
\caption{Our mask detection pipeline with data augmentation based on cycle-consistent GANs. Original training spectrograms are transferred from one class to the other using two generators, $G$ and $G'$. Original and augmented spectrograms are further used to train an ensemble of ResNet models with depths ranging from 18 layers to 101 layers. Feature vectors from the penultimate layer of each ResNet are concatenated and provided as input to an SVM classifier which makes the final prediction. Best viewed in color.}
\label{fig_pipeline}
\end{center}
\vskip -0.9cm
\end{figure*}

While deep neural networks attain state-of-the-art results in various domains \cite{He-CVPR-2016,Dahl-TASLP-2011,Georgescu-Access-2020,Krizhevsky-NIPS-2012,Long-CVPR-2015}, such models can easily succumb to the pitfall of overfitting \cite{Zhang-ICLR-2017}. This means that deep models can take decisions based on various biases existing in training data. A notorious example is an image of a wolf being correctly labeled, but only because of the snowy background \cite{Ribeiro-KDD-2016}. In our case, the training samples belonging to one class may have different gender and age distribution than the training samples belonging to the other class, among other unknown biases. Instead of finding relevant features to discriminate utterances with and without mask, a neural network might consider features for gender prediction or age estimation, which is undesired. With our data augmentation approach, all utterances with mask are translated to utterances without mask and the other way around, as shown in Figure \ref{fig_pipeline}. Any potential bias in the distribution of training data samples is eliminated through the compensation that comes with the augmented data samples from the opposite class. This forces the neural networks to discover features that discriminate the training data with respect to the desired task, i.e.~classification into \emph{mask} versus \emph{non-mask}.

We conduct experiments on the Mask Augsburg Speech Corpus (MASC), showing that our data augmentation approach attains superior results in comparison to a set of baselines, e.g. noise perturbation and time shifting, and a set of state-of-the-art data augmentation techniques, e.g.~speed perturbation \cite{Ko-INTERSPEECH-2015}, conditional GANs \cite{Chatziagapi-INTERSPEECH-2019} and SpecAugment \cite{Park-INTERSPEECH-2019}.

\section{Related Work}

Successful communication is an important component in performing effective tasks, e.g.~consider doctors in surgery rooms. While communication is crucial, doctors are often wearing surgical masks, which could lead to less effective communication. Although surgical masks affect voice clarity, human listeners reported small effects on speech understanding \cite{Mendel-JAAA-2008}. Furthermore, there is limited research addressing the effects of different face covers on voice acoustic properties. The speaker recognition task was studied in the context of wearing a face cover \cite{Saeidi-INTERSPEECH-2015,Saeidi-INTERSPEECH-2016}, but the results indicated a small accuracy degradation ratio. In addition, a negligible level of artifacts are introduced by surgical masks in automatic speech understanding \cite{Ravanelli-AISV-2013}.

To our knowledge, there are no previous works on mask detection from speech. We therefore consider augmentation methods for audio data as related work.
The superior performance of deep neural networks relies heavily on large amounts of training data \cite{LeCun-Nature-2015}. However, labeled data in many real-world applications is hard to collect. Therefore, data augmentation has been proposed as a method to generate additional training data, improving the generalization capacity of neural networks. As discussed in the recent survey of Wen et al.~\cite{Wen-ArXiv-2020}, a wide range of augmentation methods have been proposed for time series data, including speech-related tasks. A classic data augmentation method is to perturb a signal with noise in accordance to a desired signal-to-noise ratio (SNR). Other augmentation methods with proven results on speech recognition and related tasks are time shifting and speed perturbation \cite{Ko-INTERSPEECH-2015}. While these data augmentation methods are applied on raw signals, some of the most recent techniques \cite{Chatziagapi-INTERSPEECH-2019,Park-INTERSPEECH-2019} are applied on spectrograms. Representing audio signals through spectrograms goes hand in hand with the usage of CNNs or similar models on speech recognition tasks, perhaps due to their outstanding performance on image-related tasks. Park et al.~\cite{Park-INTERSPEECH-2019} performed augmentation on the log mel spectrogram through time warping or by masking blocks of frequency channels and time steps. Their experiments showed that their technique, SpecAugment, prevents overfitting and improves performance on automatic speech recognition tasks. More closely related to our work, Chatziagapi et al.~\cite{Chatziagapi-INTERSPEECH-2019} proposed to augment the training data by generating new data samples using conditional GANs \cite{Mirza-ArXiv-2014,Mariani-ArXiv-2018}. Since conditional GANs generate new data samples following the training data distribution, unwanted and unknown distribution biases in the training data can only get amplified after augmentation. Unlike Chatziagapi et al.~\cite{Chatziagapi-INTERSPEECH-2019}, we employ cycle-consistent GANs \cite{Zhu-ICCV-2017,Kim-ICLR-2020}, learning to transfer training data samples from one class to another while preserving other aspects. By transferring samples from one class to another, our data augmentation technique is able to level out any undesired distribution biases. Furthermore, we show in the experiments that our approach provides superior results.

\section{Method}

\noindent
{\bf Data representation.}
CNNs attain state-of-the-art results in computer vision \cite{He-CVPR-2016,Krizhevsky-NIPS-2012}, the convolutional operation being initially applied on images. In order to employ state-of-the-art CNNs for our task, we first transform each audio signal sample into an image-like representation. Therefore, we compute the discrete Short Time Fourier Transform (STFT), as follows:
\begin{equation}\label{eq_stft}
STFT\{x[n]\}(m, k)=\!\!\sum_{n=-\infty}^{\infty}\!\! x[n] \cdot w[n-m R] \cdot e^{-j \frac{2 \pi}{N_x}k n},
\end{equation}
where $x[n]$ is the discrete input signal, $w[n]$ is a window function (in our approach, Hamming), $N_x$ is the STFT length and $R$ is the hop (step) size \cite{Allen-IEEE-1977}. 
Prior to the transformation, we scaled the raw audio signal, dividing it by the maximum. In the experiments, we used $N_{x}\!=\!1024$, $R\!=\!64$ and a window size of $512$. We preserved the complex values (real and imaginary) of STFT and kept only one side of the spectrum, considering that the spectrum is symmetric because the raw input signal is real. Finally, each utterance is represented as a spectrogram of $2\!\times\!513\!\times\!250$ components, where $250$ is the number of time bins.

\noindent
{\bf Learning framework.}
Our learning model is based on an ensemble of residual neural networks (ResNets) \cite{He-CVPR-2016} that produce feature vectors which are subsequently joined together (concatenated) and given as input to an SVM classifier, as illustrated in Figure~\ref{fig_pipeline}. We employ ResNets because residual connections eliminate vanishing or exploding gradient problems in training very deep neural models, providing alternative pathways for the gradients during back-propagation. 
We employed four ResNet models with depths ranging from 18 to 101 layers in order to generate embeddings with different levels of abstraction. In order to combine the ResNet models, we remove the Softmax classification layers and concatenate the feature vectors (activation maps) resulting from the last remaining layers. ResNet-18 and ResNet-34 provide feature vectors of $512$ components, while ResNet-50 and ResNet-101 produce $2048$-dimensional feature vectors. After concatenation, each utterance is represented by a feature vector of $5120$ components. On top of the combined feature vectors, we train an SVM classifier. The SVM model~\cite{Cortes-ML-1995} aims at finding a hyperplane separating the training samples by a maximum margin, while including a regularization term in the objective function, controlling the degree of data fitting through the number of support vectors. We validate the regularization parameter $C$ on the development set. The SVM model relies on a kernel (similarity) function \cite{Taylor-CUP-2004,Ionescu-Springer-2016} to embed the data in a Hilbert space, in which non-linear relations are transformed into linear relations. We hereby consider the Radial Basis Function (RBF) kernel defined as
$k_{RBF}(x,y)= e^{-\gamma \lVert x - y \rVert^2}$,
where $x$ and $y$ are two feature vectors and $\gamma$ is a parameter that controls the range of possible output values. 

\begin{figure}[!t]
\begin{center}
\centerline{\includegraphics[width=1.0\columnwidth]{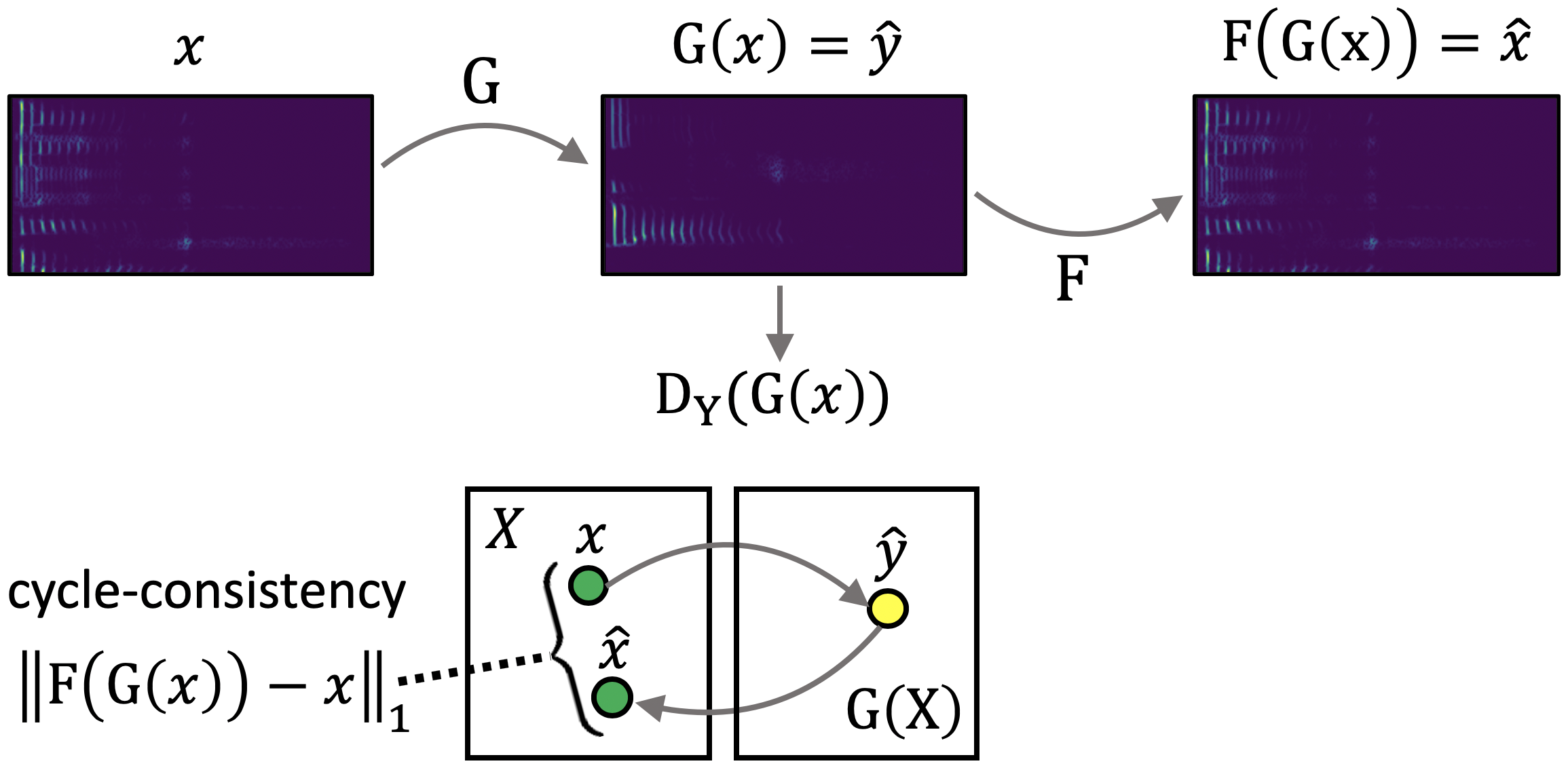}}
\vskip -0.2cm
\caption{Translating utterances (spectrograms) using cycle-consistent GANs. The spectrogram $x$ (with mask) is translated using the generator $G$ into $\hat{y}$ (without mask). The spectrogram $\hat{y}$ is translated back to the original domain $X$ through the generator $F$. The generator $G$ and the discriminator $D_Y$ are optimized in an adversarial fashion, just as in any other GAN. In addition, the GAN is optimized with respect to the cycle-consistency loss between the original spectrogram $x$ and the spectrogram $\hat{x}$. Best viewed in color.}
\label{fig_gan}
\end{center}
\vskip -0.9cm
\end{figure}

\noindent
{\bf Data augmentation.}
Our data augmentation method is inspired by the success of cycle-consistent GANs \cite{Zhu-ICCV-2017} in image-to-image translation for style transfer. Based on the assumption that style is easier to transfer than other aspects, e.g.~geometrical changes, cycle-GANs can replace the style of an image with a different style, while keeping its content. In a similar way, we assume that cycle-GANs can transfer between utterances with and without mask, while preserving other aspects of the utterances, e.g.~the spoken words are the same. We therefore propose to use cycle-GANs for utterance-to-utterance (spectrogram-to-spectrogram) transfer, as illustrated in Figure~\ref{fig_gan}.  The spectrogram $x$ (with mask) is translated using the generator $G$ into $\hat{y}$, to make it seem that $\hat{y}$ was produced by a speaker not wearing a mask. The spectrogram $\hat{y}$ is translated back to the original domain $X$ through the generator $F$. The generator $G$ is optimized to fool the discriminator $D_Y$, while the discriminator $D_Y$ is optimized to separate generated samples without mask from real samples without mask, in an adversarial fashion. In addition, the GAN is optimized with respect to the reconstruction error computed between the original spectrogram $x$ and the spectrogram $\hat{x}$. Adding the reconstruction error to the overall loss function ensures the cycle-consistency. The complete loss function of a cycle-GAN \cite{Zhu-ICCV-2017} for spectrogram-to-spectrogram translation in both directions is:
\begin{equation}\label{eq_cycle_GAN}
\begin{split}
\!\!\!L_{cycle\mbox{-}}&_{GAN} (G,F,D_X,D_Y,x,y)=L_{GAN} (G,D_Y,x,y)\\
& +L_{GAN} (F,D_X,x,y)+ \lambda \cdot L_{cycle} (G,F,x,y),
\end{split}
\end{equation}
where, $G$ and $F$ are generators, $D_X$ and $D_Y$ are discriminators, $x$ is a spectrogram from the \emph{mask} class, $y$ is a spectrogram from the \emph{non-mask} class and $\lambda$ is a parameter that controls the importance of cycle-consistency with respect to the two GAN losses. The first GAN loss is the least squares loss that corresponds to the translation from domain $X$ (with mask) to domain $Y$ (without mask):
\begin{equation}
\begin{split}
\!\!\!L_{GAN} (G,D_Y,x,y)&=E_{y \sim P_{data}(y)} \left[(D_Y (y))^2\right]\\
&+E_{x \sim P_{data} (x)} \left[(1\!-\!D_Y (G(x)))^2\right]\!,
\end{split}
\end{equation}
where $E[\cdot]$ is the expect value and $P_{data}(\cdot)$ is the probability distribution of data samples. Analogously, the second GAN loss is the least squares loss that corresponds to the translation from domain $Y$ (without mask) to domain $X$ (with mask):
\begin{equation}
\begin{split}
\!\!\!L_{GAN} (F,D_X,x,y)&=E_{x \sim P_{data}(x)} \left[(D_X (x))^2\right]\\
&+E_{y \sim P_{data} (y)} \left[(1\!-\!D_X (F(y)))^2\right]\!.
\end{split}
\end{equation}
The cycle-consistency loss in Equation~\eqref{eq_cycle_GAN} is defined as the sum of cycle-consistency losses for both translations:
\begin{equation}
\begin{split}
L_{cycle} (G,F,x,y)&=E_{x \sim P_{data} (x)} \left[ \left\lVert F(G(x))-x \right\rVert_1 \right]\\
&+E_{y \sim P_{data}(y)} \left[ \left\lVert G(F(y))-y\right\rVert_1 \right], 
\end{split}
\end{equation}
where $\left\lVert \cdot \right\rVert_1$ is the $l_1$ norm.

Although cycle-GAN is trained to simultaneously transfer spectrograms in both directions, we observed that, in practice, the second generator $F$ does not perform as well as the first generator $G$. We therefore use an independent cycle-GAN to transfer spectrograms without mask to spectrograms with mask. We denote the first generator of this cycle-GAN as $G'$. Upon training the two cycle-GANs, we keep only the generators $G$ and $G'$ for data augmentation. Hence, in the end, we are able to accurately transfer spectrograms both ways. By transferring spectrograms from one class to the other, we level out any undesired or unknown distribution biases in the training data.

In our experiments, we employ a more recent version of cycle-consistent GANs, termed U-GAT-IT \cite{Kim-ICLR-2020}. Different from cycle-GAN \cite{Zhu-ICCV-2017}, U-GAT-IT incorporates attention modules in both generators and discriminators, along with a new normalization function (Adaptive Layer-Instance Normalization), with the purpose of improving the translation from one domain to the other. The attention maps are produced by an auxiliary classifier, while the parameters of the normalization function are learned during training. Furthermore, the loss function used to optimize U-GAT-IT contains two losses in addition to those included in Equation~\eqref{eq_cycle_GAN}. The first additional loss is the sum of identity losses ensuring that the amplitude distributions of input and output spectrograms are similar:
\begin{equation}
\begin{split}
L_{identity} (G,F,x,y)&=E_{y \sim P_{data} (y)} \left[ \left\lVert G(y)-y \right\rVert_1 \right]\\
&+E_{x \sim P_{data} (x)} \left[ \left\lVert F(x)-x \right\rVert_1 \right].
\end{split}
\end{equation}
The second additional loss is the sum of the least squares losses that introduce the attention maps:
\begin{equation}
\begin{split}
\!\!\!L_{CAM} &(G,F,D_X,D_Y,x,y) = E_{y \sim P_{data} (y)} \left[(D_Y (y))^2 \right]\\
&+E_{x \sim P_{data} (x)} \left[(1-D_Y (G(x)))^2 \right]\\
&+E_{x~P_{data} (x)} \left[(D_X (x))^2 \right]\\
&+E_{y \sim P_{data} (y)} \left[(1-D_X (F(y)))^2 \right].
\end{split}
\end{equation}

\section{Experiments}

\noindent
{\bf Data set.}
The data set provided by the ComParE organizers for MSC is the Mask Augsburg Speech Corpus. The data set is partitioned into a training set of 10,895 samples, a development set of 14,647 samples and a test  set of 11,012 samples. It comprises recordings of 32 German native speakers, with or without wearing surgical masks. Each data sample (utterance) is a recording of $1$ second at a sampling rate of $16$ KHz.

\noindent
{\bf Performance measure.} The organizers decided to rank participants based on the unweighted average recall. We therefore report our performance in terms of this measure.

\noindent
{\bf Baselines.}
The ComParE organizers \cite{Schuller-INTERSPEECH-2020} provided some baseline results on the development and the private test sets. We considered their top baseline results, obtained either by a ResNet-50 model or by an SVM trained on a fusion of features. In addition, we compare our novel data augmentation method based on U-GAT-IT with several data augmentation approaches, ranging from standard approaches such as noise perturbation and time shifting to state-of-the-art methods such as speed perturbation \cite{Ko-INTERSPEECH-2015}, conditional GANs \cite{Chatziagapi-INTERSPEECH-2019} and SpecAugment \cite{Park-INTERSPEECH-2019}.


\noindent
{\bf Parameter tuning and implementation details.}
For data augmentation, we adapted U-GAT-IT \cite{Kim-ICLR-2020} in order to fit our larger input images (spectrograms). We employed the shallower architecture provided in the official U-GAT-IT code release\footnote{\scriptsize{\url{https://github.com/znxlwm/UGATIT-pytorch}}}. We adapted the number of input and output channels in accordance with our complex data representation, considering the real and the imaginary parts of the STFT as two different channels. We trained U-GAT-IT for $100$ epochs on mini-batches of $2$ samples using the Adam optimizer \cite{Kingma-ICLR-2015} with a learning rate of $10^{-4}$ and a weight decay of $10^{-4}$.
For the ResNet models, we used the official PyTorch implementation\footnote{\scriptsize{\url{https://pytorch.org/hub/pytorch_vision_resnet}}}. We only adjusted the number of input channels of the first convolutional layer, allowing us to input spectrograms with complex values instead of RGB images. We tuned the hyperparameters of the ResNet models on the development set. All models are trained for $60$ epochs with a learning rate between $10^{-3}$ and $10^{-4}$ and a mini-batch size of $16$. In order to reduce the influence of the random weight initialization on the performance, we trained each model in three trials (runs), reporting the performance corresponding to the best run. For a fair evaluation, we apply the same approach to the data augmentation baselines, i.e. we consider the best performance in three runs.
For the SVM, we experiment with the RBF kernel, setting $\gamma = 10^{-2}$. For the regularization parameter $C$ of the SVM, we consider values in the set $\{10^{-3}, 10^{-2}, 10^{-1}, 10^{0}, 10^{1}, 10^{2}, 10^{3}\}$. We tuned the regularization parameter on the development data set. For the final evaluation on the private test set, we added the development data samples to the training set.

\begin{table}[!t]
  \caption{Results of four ResNet models (ResNet-18, ResNet-34, ResNet-50, ResNet-101) in terms of unweighted average recall on the development set, with various data augmentation methods. Scores that are above the baseline without any data augmentation are highlighted in bold.}
  \label{tab_results_resnets}
\vspace{-0.2cm}
  \centering
\setlength\tabcolsep{3.0pt}
\begin{tabular}{ l c c c c}
\toprule
        \multicolumn{1}{c}{\textbf{Augmentation}} & 
        \multicolumn{4}{c}{\textbf{ResNet}}\\
        \multicolumn{1}{c}{\textbf{method}} & 
        \multicolumn{1}{c}{\textbf{18}} &
        \multicolumn{1}{c}{\textbf{34}} &
        \multicolumn{1}{c}{\textbf{50}} &
        \multicolumn{1}{c}{\textbf{101}}\\
\midrule
none                                                & 69.03         & 68.62         & 68.68         & 69.01 \\
\midrule
noise perturbation                                  & 68.37         & {\bf 69.57}   & 67.77         & 68.95 \\
time shifting                                       & {\bf 69.35}   & {\bf 69.39}   & {\bf 69.15}   & {\bf 69.42} \\
speed perturbation \cite{Ko-INTERSPEECH-2015}       & {\bf 70.14}   & 68.35         & 68.68         & 66.13\\
conditional GAN \cite{Chatziagapi-INTERSPEECH-2019} & 60.23         & 56.05         & 58.17         & 55.02\\
SpecAugment \cite{Park-INTERSPEECH-2019}            & 67.38         & {\bf 69.72}   & {\bf 69.53}   & 68.19 \\
\midrule
U-GAT-IT (ours)                                     & {\bf 69.86}   & {\bf 70.22}   & {\bf 69.88} & {\bf 70.02} \\
U-GAT-IT + time shifting (ours)                     & {\bf 71.34}   & {\bf 70.85}   & {\bf 71.16} & {\bf 70.73} \\
\bottomrule
\end{tabular}
\vspace{-0.3cm}
\end{table}

\noindent
{\bf Preliminary results.}
In Table~\ref{tab_results_resnets}, we present the results obtained by each ResNet model using various data augmentation techniques. First, we note that the augmentation based on conditional GANs~\cite{Chatziagapi-INTERSPEECH-2019} reduces the performance with respect to the baseline without data augmentation. While training the conditional GANs, we faced convergence issues, which we believe to be caused by the large size of the input spectrograms, which are more than twice as large compared to those used in the original paper~\cite{Chatziagapi-INTERSPEECH-2019}. We hereby note that GANs that learn to transfer samples~\cite{Zhu-ICCV-2017,Kim-ICLR-2020} are much easier to train than GANs that learn to generate new samples from random noise vectors~\cite{Mirza-ArXiv-2014,Mariani-ArXiv-2018}, since the transfer task is simply easier (the input is not a random noise vector, but a real data sample). While noise perturbation and speed perturbation \cite{Ko-INTERSPEECH-2015} bring performance improvements for only one of the four ResNet models, SpecAugment manages to bring improvements for two ResNet models. There are only two data augmentation methods that bring improvements for all four ResNet models. These are time shifting and U-GAT-IT. However, we observe that U-GAT-IT provides superior results compared to time shifting in each and every case. While speed perturbation brings the largest improvement for ResNet-18, our augmentation method based on U-GAT-IT brings the largest improvements for ResNet-34, ResNet-50 and ResNet-101. Among the individual augmentation methods, we conclude that U-GAT-IT attains the best results. Since time shifting and U-GAT-IT are the only augmentation methods that bring improvements for all ResNet models, we decided to combine them in order to increase our rank in the competition. We observe further performance improvements on the development set after combining U-GAT-IT with time shifting.

\begin{table}[!t]
  \caption{Results of SVM ensembles based on ResNet features, with and without data augmentation, in comparison with the official baselines \cite{Schuller-INTERSPEECH-2020}. Unweighted average recall values are provided for both the development and the private test sets.}
  \label{tab_results_final}
\vspace{-0.2cm}
  \centering
\setlength\tabcolsep{4.0pt}
\begin{tabular}{ l c c c}
    \toprule
        \multicolumn{1}{c}{\textbf{Approach}} & 
        \multicolumn{1}{c}{\textbf{SVM $\mathbf{C}$}} &
        \multicolumn{1}{c}{\textbf{Dev}} &
        \multicolumn{1}{c}{\textbf{Test}} \\
    \midrule
    DeepSpectrum \cite{Schuller-INTERSPEECH-2020}    & -        & 63.4      & 70.8 \\
    Fusion Best \cite{Schuller-INTERSPEECH-2020}     & -        & -         & 71.8 \\
    \midrule
    SVM (no augmentation)                           & $10^{-3}$ & 71.3      & 72.6 \\
    SVM + U-GAT-IT                                  & $10^{-3}$ & 72.0      & 73.5 \\
    SVM + U-GAT-IT + time shifting                  & $10^{-3}$ & 72.2      & - \\
    SVM + U-GAT-IT + time shifting                  & $10^{0}$  & 71.8      & 74.6 \\
    SVM + U-GAT-IT + time shifting                  & $10^{2}$  & 71.4         & 72.6 \\
    \bottomrule
  \end{tabular}
\vspace{-0.3cm}
\end{table}

\noindent
{\bf Submitted results.}
In Table~\ref{tab_results_final}, we present the results obtained by various ensembles based on SVM applied on concatenated ResNet feature vectors. Our SVM ensemble without data augmentation is already better than the baselines provided by the ComParE organizers \cite{Schuller-INTERSPEECH-2020}. By including the ResNet models trained with augmentation based on U-GAT-IT, we observe a performance boost of $0.9\%$ on the private test set. This confirms the effectiveness of our data augmentation approach. As time shifting seems to bring only minor improvements for the SVM, we turned our attention in another direction. Noting that the validated value of $C$ ($10^{-3}$) is likely in the underfitting zone, we tried to validate it by switching the training and the development set or by moving 5,000 samples from the development set to the training set. This generated our fourth and fifth submissions with $C=10^0$ and $C=10^2$, respectively. Our top score for MSC is $74.6\%$.

\section{Conclusion}

In this paper, we presented a system based on SVM applied on top of feature vectors concatenated from multiple ResNets. Our main contribution is a novel data augmentation approach for speech, which aims at reducing the undesired distribution bias in the training data. This is achieved by transferring data from one class to another through cycle-consistent GANs. 

\noindent
{\bf Acknowledgements.}
The research leading to these results has received funding from the EEA Grants 2014-2021, under Project contract no. EEA-RO-NO-2018-0496.

\bibliographystyle{IEEEtran}

\bibliography{mybib}

\end{document}